\documentclass[12pt]{article}
\usepackage{amssymb}
\def\be{\begin{equation}}
\def\ee{\end{equation}}
\def\bea{\begin{eqnarray}}
\def\eea{\end{eqnarray}}

\def\np#1{{\sl Nucl.~Phys.~\bf B#1}}
\def\pl#1{{\sl Phys.~Lett.~\bf B#1}}
\def\pr#1{{\sl Phys.~Rev.~\bf D#1}}
\def\prl#1{{\sl Phys.~Rev. Lett.~\bf #1}}

\def\cqg#1{{\sl Class.~Quant.~Grav.~\bf #1}}

\catcode`\@=11
\def\@citex[#1]#2{%
\if@filesw \immediate \write \@auxout {\string \citation {#2}}\fi
\@tempcntb\m@ne \let\@h@ld\relax \def\@citea{}%
\@cite{%
  \@for \@citeb:=#2\do {%
    \@ifundefined {b@\@citeb}%
      {\@h@ld\@citea\@tempcntb\m@ne{\bf ?}%
      \@warning {Citation `\@citeb ' on page \thepage \space undefined}}%
      {\@tempcnta\@tempcntb \advance\@tempcnta\@ne%
      \@tempcntb\number\csname b@\@citeb \endcsname \relax%
      \ifnum\@tempcnta=\@tempcntb 
        \ifx\@h@ld\relax%
          \edef \@h@ld{\@citea\csname b@\@citeb\endcsname}%
        \else%
          \edef\@h@ld{\ifmmode{-}\else--\fi\csname b@\@citeb\endcsname}%
        \fi%
      \else
        \@h@ld\@citea\csname b@\@citeb \endcsname%
        \let\@h@ld\relax%
      \fi}%
    \def\@citea{,\penalty\@highpenalty\,}%
  }\@h@ld
}{#1}}

\def\@citeb#1#2{{[#1]\if@tempswa , #2\fi}}
\def\@citeu#1#2{{$^{#1}$\if@tempswa , #2\fi }}
\def\@citep#1#2{{#1\if@tempswa , #2\fi}}

\def\bcites{         
        \catcode`\@=11
        \let\@cite=\@citeb
        \catcode`\@=12
}

\def\upcites{         
        \catcode`\@=11
        \let\@cite=\@citeu
        \catcode`\@=12
}

\def\plaincites{      
        \catcode`\@=11
        \let\@cite=\@citep
        \catcode`\@=12
}

\newcount\hour
\newcount\minute
\newtoks\amorpm
\hour=\time\divide\hour by 60
\minute=\time{\multiply\hour by 60 \global\advance\minute by-\hour}
\edef\standardtime{{\ifnum\hour<12 \global\amorpm={am}%
        \else\global\amorpm={pm}\advance\hour by-12 \fi
        \ifnum\hour=0 \hour=12 \fi
        \number\hour:\ifnum\minute<10 0\fi\number\minute\the\amorpm}}
\edef\militarytime{\number\hour:\ifnum\minute<10 0\fi\number\minute}

\def\draftlabel#1{{\@bsphack\if@filesw {\let\thepage\relax
   \xdef\@gtempa{\write\@auxout{\string
      \newlabel{#1}{{\@currentlabel}{\thepage}}}}}\@gtempa
   \if@nobreak \ifvmode\nobreak\fi\fi\fi\@esphack}
        \gdef\@eqnlabel{#1}}
\def\@eqnlabel{}
\def\@vacuum{}
\def\marginnote#1{}
\def\draftmarginnote#1{\marginpar{\raggedright\scriptsize\tt#1}}
\def\draft{
        \pagestyle{plain}
        \overfullrule=2pt
        \oddsidemargin -.75truein
        \def\@oddhead{\sl \phantom{\today\quad\militarytime} \hfil
        \smash{\Large\sl DRAFT} \hfil \today\quad\militarytime}
        \let\@evenhead\@oddhead
        \let\label=\draftlabel
        \let\marginnote=\draftmarginnote
        \def\ps@empty{\let\@mkboth\@gobbletwo
        \def\@oddfoot{\hfil \smash{\Large\sl DRAFT} \hfil}
        \let\@evenfoot\@oddhead}
        \def\@eqnnum{(\theequation)\rlap{\kern\marginparsep\tt\@eqnlabel}%
        \global\let\@eqnlabel\@vacuum}  }

\begin{document}


\hfill UTHET-03-0601

\vspace{-0.2cm}

\begin{center}
\Large
{\bf Perturbative calculation of quasi-normal modes of Schwarzschild black holes}
\normalsize

\vspace{0.8cm}
{\bf
Suphot Musiri}\footnote{email: suphot@swu.ac.th} \\
Department of Physics, \\
Srinakharinwiroth University, Bangkok, \\
Thailand.
\vspace{.5cm}

{\bf George Siopsis}\footnote{email: gsiopsis@utk.edu}\\ Department of Physics
and Astronomy, \\
The University of Tennessee, Knoxville, \\
TN 37996 - 1200, USA.

\end{center}

\vspace{0.8cm}
\large
\centerline{\bf Abstract}
\normalsize
\vspace{.5cm}

We discuss a systematic method of analytically calculating the
asymptotic form of quasi-normal frequencies of a four-dimensional
Schwarzschild black hole by expanding around the zeroth-order approximation
to the wave equation proposed by Motl and Neitzke.
We obtain an explicit expression for the first-order correction and arbitrary
spin. Our results are in
agreement with the results from WKB and numerical analyses in the case of
scalar and gravitational waves.

\newpage


The study of quasi-normal modes in asymptotically flat space-times has attracted
a lot of attention recently~\cite{bibx1,bibx2,bibx3,bibx4,bibx5,bibx6,bibx7,bibx8,bibx9,bibx10,bibx11} because the asymptotic form of quasi-normal
frequencies was shown to be related to the Barbero-Immirzi parameter~\cite{bibbi1,bibbi2} of Loop Quantum Gravity (see \cite{biblqg1,biblqg2,biblqg3,biblqg4,biblqg5} and references therein).
Quasi-normal modes form a discrete spectrum of complex frequencies whose imaginary part determines the decay rate of the fluctuation.
The asymptotic form of high overtones is
\be\label{eq1} \frac{\omega_n}{T_H} = (2n+1)\pi i + \ln 3 \ee
where $T_H$ is the Hawking temperature.
This has been derived numerically~\cite{bibn1,bibn2,bibn3,bibn4,bibn5}
and subsequently confirmed analytically~\cite{bibx3,bibx5}.
The imaginary part is large making the numerical analysis cumbersome, but
is easy to understand by noting that the spacing of the frequencies $2\pi i T_H$
coincides with the spacing of the poles of a thermal Green function on the
Schwarzschild black hole background.
The analytical value of the real part was first conjectured by Hod~\cite{bibhod}
based on the general form of the area spectrum of black holes
proposed by Mukhanov and Bekenstein~\cite{bibmb} ($\delta A =4G \ln k$,
$k = 2,3,\dots$, where $\delta A$ is the spacing of eigenvalues and
$G$ is Newton's constant in units such that $\hbar = c = 1$)
which is related to the number of microstates of the black hole through
the entropy-area relation $S = \frac{1}{4G}\ A$~\cite{bibbeke}. It has an intriguing value
from the loop quantum gravity point of view suggesting that the gauge group
should be $SO(3)$ rather than $SU(2)$ (since we have $k=3$ instead of $k=2$). Thus the study of quasi-normal modes
may lead to a deeper understanding of black holes and quantum gravity.

The analytic derivation of the asymptotic form of quasi-normal frequencies~(\ref{eq1}) by Motl and Neitzke~\cite{bibx5} offered a new surprise as it heavily
relied on the black hole singularity. It is intriguing that the ``unphysical''
region beyond the horizon influences the behavior of physical quantities.
Here we extend the results of ref.~\cite{bibx5} by calculating the first-order
correction to the asymptotic formula~(\ref{eq1}). We solve the wave equation
perturbatively for arbitrary spin of the wave and obtain explicit
expressions for the quasi-normal frequencies. Our results are in agreement with
numerical results~\cite{bibn3,bibx6} in the case of gravitational and scalar
waves as well as results
from a WKB analysis~\cite{bibx8} in the case of gravitational waves.

The metric of a four-dimensional Schawrzschild black hole may be written as
\be
ds^2 = -h(r)\, dt^2 + \frac{dr^2}{ h(r)} +r^2 d \Omega^2
\ \ , \ \
h(r) = 1 - \frac{2G M}{r}\ee
where $G$ is Newton's constant and $M$ is the mass of the black hole.
The Hawking temperature is
\be\label{eqth}
T_H = \frac{1}{8\pi GM} = \frac{1}{4\pi r_0}
\ee
where $r_0 = 2GM$ is the radius of the horizon.
A spin-$j$ perturbation of frequency $\omega$ is governed by the radial equation
\be\label{eq5a}
-h(r) \frac{d}{d r} \left( h(r)\frac{d\Psi}{d r}
\right) + V(r) \Psi  =\omega^2 \Psi \ee
where $V(r)$ is the ``Regge-Wheeler'' potential~\cite{bibV1,bibV2}
\be\label{eqV} V(r) = h(r) \left( \frac{\hat L^2}{r^2} + \frac{(1-j^2)r_0}{r^3} \right)\ee
and $\hat L = \ell (\ell +1)$ is the angular momentum operator.
We have $j=0,1/2,1,3/2,2$ for a scalar, Dirac, electromagnetic, gravitino
and gravitational wave, respectively. We shall keep the discussion general.
In fact, it will be necessary to avoid integer values of $j$ throughout
the discussion and only take the limit $j\to$ integer at the end of the
calculation.

It is convenient to express eq.~(\ref{eq5a}) in terms of the ``tortoise coordinate''
\be r_* = r + r_0 \ln \left(\frac{r}{r_0} -1 \right) \ee
We obtain
\be\label{eq5b}
-\frac{d^2\Psi}{d r_*^2} + V(r(r_*)) \Psi  =\omega^2 \Psi \ee
to be solved along the entire real axis.
At both ends the potential vanishes ($V\to 0$ as $r_*\to \pm\infty$)
so the solutions behave as $\Psi\sim e^{\pm i\omega r_*}$.
For quasi-normal modes we demand
\be \Psi \sim e^{\mp i\omega r_*} \ \ , \ \ r_* \to \pm\infty\ee
assuming $\mathrm{Re} \omega > 0$.
Let us express the quasi-normal mode as
\be \Psi = e^{-i\omega r_*} f(r_*) \ee
Then $f(r_*)\sim 1$ as $r_*\to +\infty$ and $f(r_*)\sim e^{2i\omega r_*}$
as $r_*\to -\infty$ (near the horizon).
Following ref.~\cite{bibx5}, we continue the coordinate $r$ analytically into
the complex plane and define the requisite boundary condition at the horizon
in terms of the monodromy
of $f(r_*(r))$ around the singular point $r=r_0$. It is easily seen to be given by
\be\label{eqmono} \mathcal{M} (r_0) = e^{-4\pi\omega r_0}\ee
along a contour running counterclockwise. This will serve as the definition of
the boundary condition at the horizon.
The contour surrounding the singularity $r=r_0$ can be deformed in the
complex $r$-plane so that it either lies beyond the horizon ($\mathrm{Re} r
< r_0$) or at infinity ($r\to \infty$).
Then the monodromy only gets a contribution from the segment lying beyond the
horizon. It is convenient to change variables to
\be z = \omega (r_* -i\pi r_0) = \omega (r + r_0 \ln (1-r/r_0))\ee
where we choose the branch that yields $z\to 0$ as $r\to 0$.
The potential can be written as a series in $\sqrt{z}$,
\be\label{eqVa} V(z) = -\frac{\omega^2}{4z^2} \left( 1-j^2 + \frac{3\ell (\ell+1) +1- j^2 }{3}\, \sqrt{-\frac{2z}{\omega r_0}} + \dots \right)\ee
which is also a formal expansion in $1/\sqrt\omega$.

To compute the monodromy around the singularity $r=r_0$, we shall deform the
contour so that it gets mapped onto the real axis in the $z$-plane. Near the
singularity $z=0$, we have $z \approx -\frac{\omega}{2r_0}\, r^2$.
We shall choose our contour in the complex $r$ plane so that near $r=0$,
the positive real axis and the negative real axis in the $z$-plane are
mapped onto
\be \arg r = \pi - \frac{\arg\omega}{2} \ \ , \ \ \arg r = \frac{3\pi}{2}
- \frac{\arg\omega}{2} \ee
in the $r$-plane.
These two segments form a $\pi /2$ angle (independent of $\arg\omega$). To avoid the $r=0$ singularity,
we shall go around an arc of angle $3\pi /2$. This translates to an angle
$3\pi$ around $z=0$ in the $z$-plane.

By considering the black hole singularity ($r=0$), we obtain
two linearly independent solutions. They are of the form
\be\label{eqZ} f_\pm (r) = r^{1\pm j} Z_\pm (r) \ee
where $Z_\pm$ are analytic functions of $r$.
By going around an arc of angle of $3\pi/2$, these change to
\be\label{eq3pi} f_\pm (e^{3\pi i/2} r) = e^{3\pi (1\pm j) i/2} \, f_\pm (r)\ee
This is an exact result which follows directly from the wave equation~(\ref{eq5b}).
To proceed further, we need to relate the behavior near the black hole singularity
to the behavior at large values of $r$ in the complex $r$-plane.
We shall do this by solving the wave equation~(\ref{eq5b}) perturbatively.
This will enable us to write the wavefunction as a perturbation series in
$1/\sqrt\omega$.

To zeroth order in $1/\sqrt\omega$, eq.~(\ref{eq5b}) can be written as
\be\label{eq5bb}
\frac{d^2\Psi^{(0)}}{d z^2} + \left( \frac{1-j^2}{4z^2} + 1 \right) \Psi^{(0)}  = 0 \ee
where we replaced the potential $V$ (eq.~(\ref{eqV})) by its zeroth-order
approximation (see eq.~(\ref{eqVa})), as suggested in ref.~\cite{bibx5}.
The solutions are given in terms of Bessel functions. They can be
chosen to match the two exact solutions~(\ref{eqZ}), respectively.
We obtain
\be\label{eq17} f_\pm^{(0)} (z) = e^{iz}\, \Psi_\pm^{(0)} = e^{iz} \sqrt{\frac{\pi z}{2}} J_{\pm j/2} (z)\ee
whose behavior at infinity ($z\to\infty$) is
\be\label{eqff0} f_\pm^{(0)} (z) \sim e^{iz} \cos (z - \pi (1\pm j)/4)\ee
By our boundary condition, we need $f(z)$ to approach a constant as $z\to\infty$
along the positive real axis in the $z$-plane.
This is accomplished by adopting the linear combination
\be f^{(0)} = f_+^{(0)} - e^{-\pi ji/2}\, f_-^{(0)} \sim e^{iz} \sqrt z\,
H_{j/2}^{(1)} (z)\ee
which approaches a constant at infinity,
\be\label{eqm1} f^{(0)}(z) \sim - e^{-\pi (1+j) i/4} \sin (\pi j/2) \ee
on account of eq.~(\ref{eqff0}).
Going along the $3\pi $ arc around $z=0$ in the $z$-plane, we obtain
\be f^{(0)}(e^{3\pi i} z) = e^{3\pi (1+j)i/2} \left( f_+^{(0)} (z) - e^{-7\pi ji/2} \, f_-^{(0)} (z)\right) \ee
whose behavior at infinity is found to be
\be\label{eqm2} f^{(0)}(z) \sim e^{-\pi (1+j) i/4} \sin (3\pi j/2) + e^{\pi (1-j)i/4} \sin (2\pi j) e^{2iz} \ee
using eq.~(\ref{eqff0}) again.
The monodromy~(\ref{eqmono}) to zeroth order is deduced from eqs.~(\ref{eqm1})
and (\ref{eqm2}) to be
\be \mathcal{M} (r_0) = -\frac{\sin (3\pi j/2)}{\sin (\pi j/2)}
= - (1+ 2\cos (\pi j)) \ee
which yields a discrete set of complex frequencies (quasi-normal modes),
\be\label{eqf0} \frac{\omega_n}{T_H} = (2n+1)\pi i + \ln (1+2\cos(\pi j) ) + o(1/\sqrt n) \ee
This reduces to the expression~(\ref{eq1}) originally obtained numerically in the case of even integer $j$ (scalar and gravitational perturbations).
It was derived analytically in refs.~\cite{bibx3,bibx5}.

To go beyond the zeroth order, expand the wavefunction in $1/\sqrt\omega$,
\be \Psi = \Psi^{(0)} + \frac{1}{\sqrt{-\omega r_0}}\, \Psi^{(1)} + o(1/\omega)\ee
The first-order correction obeys the equation
\be\label{eq5ba}
\frac{d^2\Psi^{(1)}}{d z^2} + \left( \frac{1-j^2}{4z^2} + 1 \right) \Psi^{(1)}  =\sqrt{-\omega r_0}\, \delta V \Psi^{(0)}
\ \ , \ \ \delta V(z) = \frac{1-j^2}{4z^2} + \frac{1}{\omega^2}\, V(r(z)) \ee
as is easily deduced from eqs.~(\ref{eq5b}), (\ref{eqVa}) and (\ref{eq5bb}).
The solutions of the first-order wave equation (\ref{eq5ba}) may be written in
terms of the solutions of the zeroth-order approximation~(\ref{eq5bb}) in a
standard fashion,
\be \Psi_\pm^{(1)} (z) = \mathcal{C} \Psi_+^{(0)} (z)\int_0^z \Psi_-^{(0)}
\delta V \Psi_\pm^{(0)} - \mathcal{C} \Psi_-^{(0)} (z)\int_0^z \Psi_+^{(0)}
\delta V \Psi_\pm^{(0)}
\ \ , \ \ \mathcal{C} = \frac{\sqrt{-\omega r_0}}{\sin (\pi j/2)}
\ee
where we are integrating along the positive real axis on the $z$-plane ($z>0$).
The large-$z$ behavior is found to be
\be\label{eq28} \Psi_\pm^{(1)} (z) \sim c_{-\pm} \cos (z-\pi (1+ j)/4) - c_{+\pm} \cos (z-\pi (1-j)/4)
\ \ , \ \ c_{\pm\pm} = \mathcal{C} \int_0^\infty \Psi_\pm^{(0)} \delta V \Psi_\pm^{(0)}
\ee
To find the small-$z$ behavior, we expand
\be \delta V(z) = -\frac{3\ell (\ell+1) + 1-j^2 }{6\sqrt{-2\omega r_0} }\, z^{-3/2} + o(1/\omega) \ee
where we used eqs.~(\ref{eqVa}) and (\ref{eq5ba}).
It follows that
\be\label{eq3pia} \Psi_\pm^{(1)} = z^{1\pm j/2} G_\pm (z) + o(1/\omega) \ee
where $G_\pm$ are even analytic functions of $z$.
For the desired behavior at infinity, define
\bea\label{eqpsi1} \Psi &=& \Psi_+^{(0)} + \frac{1}{\sqrt{-\omega r_0}}\, \Psi_+^{(1)} - e^{-\pi ji/2} \left( 1- \frac{1}{\sqrt{-\omega r_0}}\, \xi \right)
\left(\Psi_-^{(0)} + \frac{1}{\sqrt{-\omega r_0}}\, \Psi_-^{(1)}\right) + \dots\nonumber \\
&=& \Psi^{(0)} + \frac{1}{\sqrt{-\omega r_0}}\, \left\{ \Psi_+^{(1)} - e^{-\pi ji/2} \Psi_-^{(1)} + e^{-\pi ji/2} \xi\Psi_-^{(0)}\right\}
+ \dots \eea
where $\xi  \sim o(1)$ and the dots represent terms of order higher than $o(1/\sqrt\omega)$.
Demanding $\Psi\sim e^{-iz}$ as $z\to +\infty$ fixes the parameter $\xi$ to
\be\label{eq32} \xi =  \xi_+ + \xi_- \ \ , \ \
\xi_+ = c_{++} e^{\pi ji/2} - c_{+-}\ \ , \ \
\xi_- = c_{--} e^{-\pi ji/2} - c_{+-}\ee
Then $f = e^{iz} \Psi$ approaches a constant,
\be f(z) \sim - e^{-\pi (1+j)i/4} \sin (\pi j/2) \left\{ 1 - \frac{\xi_-}{\sqrt{-\omega r_0}} \right\} \ee
as $z\to +\infty$ along the real axis.
Turning to values of $z$ in the neighborhood of the black hole singularity
(around $z=0$), we observe that
by going around a $3\pi$ arc, we have
\be \Psi_\pm^{(1)} (e^{3\pi i} z) = e^{3\pi (2\pm j) i/2 } \Psi_\pm^{(1)} (z) \ee
Therefore, eqs.~(\ref{eq3pi}), (\ref{eq3pia}) and (\ref{eqpsi1}) imply
\be \Psi (e^{3\pi i} z) = \Psi^{(0)} (e^{3\pi i} z) -i e^{3\pi (1+j)i/2}
\frac{1}{\sqrt{-\omega r_0}}\,
\left\{ \Psi_+^{(1)} (z) - e^{-7\pi ji/2} (\Psi_-^{(1)} (z) - i\xi \Psi_-^{(0)}
(z)) \right\} \ee
As $z\to -\infty$ along the real axis, we deduce
\be f (z) \sim e^{-\pi (1+j)i/4} \sin (3\pi j/2) \left\{ 1-\frac{1}{\sqrt{-\omega r_0}}\, A\right\} + e^{\pi (1-j)i/4}
\sin(2\pi j) \left\{ 1-\frac{1}{\sqrt{-\omega r_0}}\, B\right\} e^{2iz}\ee
where
\be A = \frac{1-i}{2}\ e^{\pi ji/2}\ \left( \xi_++i\xi_- -  \xi \, \cot (3\pi j/2)\right)\ee
and $B$ is a constant that can be easily calculated but is not needed in our
discussion.
The monodromy~(\ref{eqmono}) to this order is
\be \mathcal{M} (r_0) = - \frac{\sin (3\pi j/2)}{\sin (\pi j/2)}
\, \left\{ 1 + \frac{i-1}{2\sqrt{-\omega r_0}}\ e^{\pi ji/2}\ \left( \xi_- -\xi_+
+ \xi \cot(3\pi j/2) \right) \right\} \ee
leading to the quasi-normal frequencies
\be\label{eqf1} \frac{\omega_n}{T_H} = (2n+1)\pi i + \ln (1+2\cos(\pi j) ) + \frac{e^{\pi ij/2}}{\sqrt{n+1/2}} \left( \xi_- -\xi_+
+ \xi \cot(3\pi j/2) \right) + o(1/n) \ee
which includes a $o(1/\sqrt n)$ correction to the $o(1)$ asymptotic formula~(\ref{eqf0}).
To obtain an explicit expression, we need the integral
\be \mathcal{J} (\nu,\mu) \equiv \int_0^\infty dz\, z^{-1/2} J_\nu (z) J_\mu (z) = \frac{\sqrt{\pi/2}
\, \Gamma (\frac{\nu+\mu + 1/2}{2})}{\Gamma (\frac{-\nu+\mu+3/2}{2})
\Gamma (\frac{\nu+\mu+3/2}{2}) \Gamma (\frac{\nu-\mu +3/2}{2} )}\ee
On account of~(\ref{eq17}),  we obtain
\be c_{\pm\pm} = \pi\, \frac{3\ell (\ell + 1) +1-j^2}{12\sqrt 2\, \sin (\pi j/2)}
\, \mathcal{J} (\pm j/2, \pm j/2)\ee
for the coefficients in the first-order wavefunction correction~(\ref{eq28}).
After some algebra, we obtain from eq.~(\ref{eq32}) an explicit expression
for the combination of constants needed in the $o(1/\sqrt n)$ contribution to
the quasi-normal frequencies (eq.~(\ref{eqf1})),
\bea\label{eqxi}  \xi_- -\xi_+
+ \xi \cot(3\pi j/2) &=& (1-i) \ \frac{3\ell(\ell +1) +1-j^2}{24\sqrt 2 \pi^{3/2}}
\ \frac{\sin (2\pi j)}{\sin (3\pi j/2)}\nonumber \\
& & \times \Gamma^2 (1/4)\ \Gamma(1/4 + j/2)
\ \Gamma (1/4 - j/2)\eea
where we used the identity 
\be \Gamma (y) \Gamma (1-y) = \frac{\pi}{\sin (\pi y)} \ee
Eq.~(\ref{eqxi}) has a well-defined finite limit as $j$ approaches an integer.

In the limit $j\to 0^+$ (scalar perturbation), eq.~(\ref{eqf1}) reduces to
\be\label{eqfsc} \frac{\omega_n}{T_H} = (2n+1)\pi i + \ln 3 + \frac{1-i}{\sqrt{n+1/2}} \ \frac{\ell(\ell +1) +1/3}{6\sqrt 2 \pi^{3/2}}
\ \Gamma^4 (1/4) + o(1/n) \ee
in agreement with numerical results~\cite{bibx6}.

For gravitational waves ($j\to 2$), we obtain
\be\label{eqfgr} \frac{\omega_n}{T_H} = (2n+1)\pi i + \ln 3 + \frac{1-i}{\sqrt{n+1/2}} \ \frac{\ell(\ell +1) -1}{18\sqrt 2 \pi^{3/2}}
\ \Gamma^4 (1/4) + o(1/n) \ee
in agreement with the result from a WKB analysis~\cite{bibx8} as well as numerical results~\cite{bibn3}. Explicit expressions for different spins may be readily deduced from the general expression~(\ref{eqf1}) using (\ref{eqxi}).

To summarize, we have derived analytically the first-order correction to the asymptotic
form of quasi-normal frequencies~(\ref{eq1}) for four-dimensional Schwarzschild
black holes.
We solved the wave equation perturbatively based on the zeroth-order solution
proposed by Motl and Neitzke~\cite{bibx5} and obtained an explicit expression
for arbitrary spin of the wave (eqs.~(\ref{eqf1}) and (\ref{eqxi})). Our results agreed with numerical results~\cite{bibn3,bibx6} in the case of scalar and gravitational waves, as well as the analytical expression found by WKB analysis~\cite{bibx8}
in the case of gravitational waves.
It would be interesting to extend our method to more general space-time backgrounds such as Kerr black holes~\cite{bibx10} and higher dimensions~\cite{bibx11}.

\section*{Acknowledgments}

G.~S.~is supported by the US Department of Energy under grant
DE-FG05-91ER40627.

\end{document}